\documentclass[%
preprint,
 amsmath,amssymb,
 aps,
 longbibliography,
]{revtex4-1}

\usepackage{graphicx}
\usepackage{dcolumn}
\usepackage{bm}
\usepackage{url}
\usepackage[dvipdfmx, colorlinks=true, citecolor=magenta, linkcolor=blue]{hyperref}%
\usepackage{enumerate}

\begin{document}

\title{Neutrino mixing and CP violation from Dirac-Majorana mixing}%

\author{Junpei Harada}
 \email{jharada@hoku-iryo-u.ac.jp}
\affiliation{%
 Research Center for Higher Education, Health Sciences University of Hokkaido, Japan
}%

\date{December 21, 2015}

\begin{abstract}
We consider a minimal condition that predicts the 1-3 lepton mixing angle $\theta_{13} \simeq \theta_{\rm C}/\sqrt{2}$ with $\theta_{\rm C}$ the Cabibbo angle, and give the improved prediction of $\theta_{13}$. In the case of normal mass ordering, the theoretical value of $\theta_{13}$ is predicted as $\theta_{13}=8.6^\circ$, which is in good agreement with the current global best fit. In the case of inverted mass ordering, the theoretical value is predicted as $\theta_{13}=9.7^\circ$, which is far from the current global best fit. We also study the leptonic CP violation. We show that any values of the leptonic Dirac CP phase $\delta_{\rm CP}$ can be obtained by fine tuning. Without fine tuning, $|\sin\delta_{\rm CP}|$ should be very small, typically of ${\cal O}(\lambda^2)$ where $\lambda\equiv \sin\theta_{\rm C}$. Furthermore, a model-independent measure of CP violation is proposed, which is applicable to any flavor models. 
\end{abstract}


\maketitle
In 2012, relatively large value of the lepton mixing angle, $\theta_{13} \simeq 9^\circ$, has been established by the neutrino oscillation experiments,  T2K~\cite{Abe:2011sj}, MINOS~\cite{Adamson:2011qu}, Double-Chooz~\cite{Abe:2011fz}, Daya-Bay~\cite{An:2012eh}, and RENO~\cite{Ahn:2012nd}. The discovery of $\theta_{13}\simeq 9^\circ$ is indeed an exciting progress in neutrino physics. Experimentally, the determination of the leptonic Dirac CP phase $\delta_{\rm CP}$, neutrino mass ordering (normal or inverted), and the octant of the 2-3 lepton mixing angle $\theta_{23}$ ($\theta_{23} < \pi/4$ or $\theta_{23} > \pi/4$), is important in neutrino oscillations. The current status and next generation of neutrino studies are summarized in refs.~\cite{Altarelli:2014dca, Gonzalez-Garcia:2014bfa, Smirnov:2015rsa}. 

Theoretically, it is interesting that $\theta_{13} \simeq 9^\circ$ was predicted before the determination~\cite{Minakata:2004xt,Harada:2005km,Chauhan:2006im}. When $\theta_{13} \simeq 9^\circ$ was predicted, the best fit value of global data was given by $\theta_{13}=0^\circ$~\cite{Maltoni:2004ei}. Although the agreement between theoretical prediction and the measured value is important, it makes theoretical prediction interesting that the prediction was given by $\theta_{13} \simeq \theta_{\rm C}/\sqrt{2}$ with $\theta_{\rm C}$ the Cabibbo angle, which indicates that the lepton flavor mixing and the quark flavor mixing are correlated. If this correlation---a nontrivial link between leptons and quarks---is true, it may play an important role to understand the flavor mixing. 

In this paper we consider a minimal condition that predicts $\theta_{13} \simeq \theta_{\rm C}/\sqrt{2}$, and give the improved prediction of $\theta_{13}$. In the case of normal mass ordering, theoretical value of $\theta_{13}$ is predicted as $\theta_{13} = 8.6^\circ$, which is consistent with the current global best fit. In the case of inverted mass ordering, theoretical value is predicted as $\theta_{13}=9.7^\circ$, which is far from the current global best fit. We also study the leptonic CP violation. We show that any values of the leptonic Dirac CP phase $\delta_{\rm CP}$ can be obtained by fine tuning, including $\delta_{\rm CP}\simeq 3\pi/2$ which has been reported by T2K~\cite{Abe:2015awa} and NO$\nu$A~\cite{Bian:2015opa} as the preferred value. Without fine tuning, $|\sin\delta_{\rm CP}|$ should be very small, typically of ${\cal O}(\lambda^2)$ where $\lambda\equiv \sin\theta_{\rm C}$. Finally, we propose a model-independent measure of CP violation, which is applicable to any flavor models. We do not consider Majorana phases in this paper, which are irrelevant for neutrino oscillations. 

The interesting relation, $\theta_{13} \simeq \theta_{\rm C}/\sqrt{2}$, is obtained as follows. The crucial condition is 
\begin{align}
	\nu_e^\prime \perp \nu_3, \label{eq:nu_e_nu3}
\end{align}
where $\nu_e^\prime$ is the Dirac mass eigenstate of the left-handed electron neutrino, and $\nu_3$ is the light mass eigenstate with mass $m_3$. Eq.~\eqref{eq:nu_e_nu3} can be exact. In that case, symmetries may exist behind eq.~\eqref{eq:nu_e_nu3}. Some deviations from eq.~\eqref{eq:nu_e_nu3} are also possible. We consider both small deviations and large deviations later. In the following we first consider the case that eq.~\eqref{eq:nu_e_nu3} is exact. 

The lepton flavor mixing is written by
\begin{align}
	\nu_\alpha = U_{\rm PMNS} \nu_i, \label{eq:PMNS}
\end{align}
where $\nu_\alpha$ $(\alpha=e, \mu, \tau)$ is the flavor eigenstate, $\nu_i$ $(i=1, 2, 3)$ is the light mass eigenstate, and $U_{\rm PMNS}$ is the Pontecorvo-Maki-Nakagawa-Sakata (PMNS) lepton mixing matrix. The PMNS matrix $U_{\rm PMNS}$ contains three mixing angles, the leptonic Dirac CP phase $\delta_{\rm CP}$, and two Majorana CP phases (in this paper we do not consider them). One can also construct the basis of the Dirac mass eigenstate for left-handed neutrinos, written by $\nu_\alpha^\prime$ $(\alpha=e, \mu, \tau)$. In that basis, the Dirac neutrino mass matrix is diagonal. Each Dirac mass eigenstate $\nu_\alpha^\prime$ can be written as a superposition of the flavor eigenstate $\nu_\alpha$. Although the weight is not determined from the Standard Model, we write $\nu_e^\prime$ as follows,
\begin{align}
	\nu_e^\prime = V_{ud} \nu_e + V_{us} \nu_\mu + V_{ub} \nu_\tau, \label{eq:nu_e_nu_alpha}
\end{align}
where $V_{ud}$, $V_{us}$, and $V_{ub}$ are the matrix elements of the Cabibbo-Kobayashi-Maskawa (CKM) quark flavor mixing matrix. Eq.~\eqref{eq:nu_e_nu_alpha} is naturally expected in $SO(10)$ or $E_6$ grand unification, since the Dirac mass matrix is approximately equivalent between quark and lepton sector. 

From eqs.~\eqref{eq:PMNS} and~\eqref{eq:nu_e_nu_alpha}, $\nu_e^\prime$  can be written as a superposition of $\nu_1, \nu_2$ and $\nu_3$,
\begin{align}
	\nu_e^\prime =& (V_{ud} U_{e1} + V_{us} U_{\mu 1} + V_{ub} U_{\tau 1})\nu_1 \nonumber \\
	& + (V_{ud} U_{e2} + V_{us} U_{\mu 2} + V_{ub} U_{\tau 2})\nu_2
	  + (V_{ud} U_{e3} + V_{us} U_{\mu 3} + V_{ub} U_{\tau 3})\nu_3, \label{eq:nu_eD_nui}
\end{align}
where $U_{\alpha i}$ $(\alpha=e, \mu, \tau, i=1, 2, 3)$ are the matrix elements of the PMNS matrix. 
In eq.~\eqref{eq:nu_eD_nui}, the coefficient of $\nu_3$ vanishes, 
\begin{align}
	V_{ud} U_{e3} + V_{us} U_{\mu 3} + V_{ub} U_{\tau 3} = 0, \label{eq:CKM-PMNS}
\end{align}
since $\nu_e^\prime$ is orthogonal to $\nu_3$. Eq.~\eqref{eq:CKM-PMNS} contains 7 physical parameters---three quark mixing angles, the Kobayashi-Maskawa quark CP phase, two lepton mixing angles, $\theta_{13}$ and $\theta_{23}$, and the leptonic Dirac CP violation phase, $\delta_{\rm CP}$. The leptonic 1-2 mixing angle, $\theta_{12}$, is irrelevant. 

		\begin{figure}[h]
			\begin{center}
				\includegraphics[width=150mm]{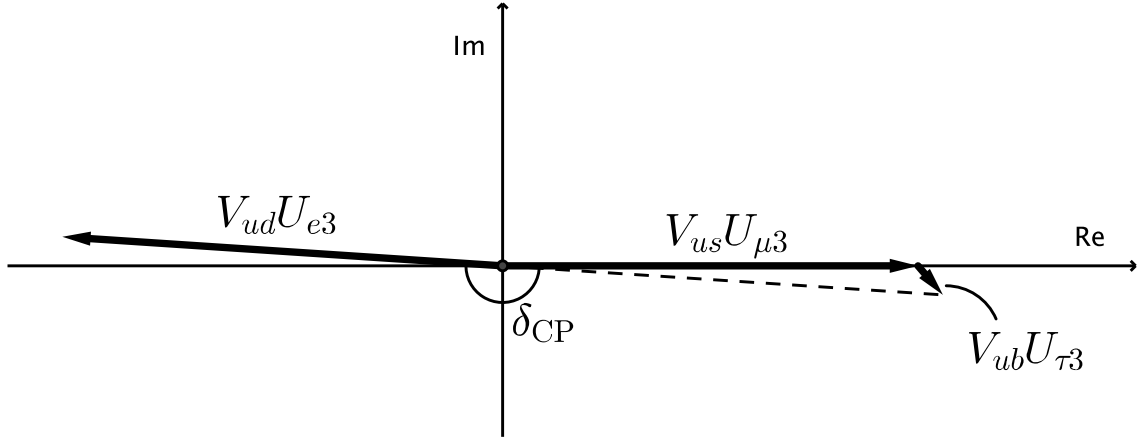}
				\caption{\label{fig:triangle}Schematic figure of eq.~\eqref{eq:CKM-PMNS}. The $|V_{ub} U_{\tau 3}|$ is much smaller than the $|V_{ud} U_{e3}|$ or $|V_{us} U_{\mu 3}|$. In the $|V_{ub} U_{\tau 3}| \rightarrow 0$ limit, $|V_{ud} U_{e3}|(\simeq \sin\theta_{13})$ equals to $|V_{us} U_{\mu 3}|$. It gives $\theta_{13} \simeq \theta_{\rm C}/\sqrt{2}$. }
			\end{center}
		\end{figure}					

Fig.~\ref{fig:triangle} is a schematic figure of eq.~\eqref{eq:CKM-PMNS}. As shown in fig.~\ref{fig:triangle}, $|V_{ub}U_{\tau 3}|$ is much smaller than $|V_{ud} U_{e3}|$ or $|V_{us} U_{\mu 3}|$. Therefore, at the lowest order approximation (or equivalently $|V_{ub}U_{\tau 3}| \rightarrow 0$ limit),  $|V_{ud}U_{e3}|$ equals to $|V_{us}U_{\mu 3}|$. Furthermore, $|V_{ud}U_{e3}|$ is typically equivalent to $\sin\theta_{13}$, since $V_{ud}\simeq 1$. Consequently, the value of $\sin\theta_{13}$ approximately equals to $|V_{us}U_{\mu 3}|$. It gives $\theta_{13} \simeq \theta_{\rm C}/\sqrt{2}$. This is a minimal overview of the interesting relation, $\theta_{13} \simeq \theta_{\rm C}/\sqrt{2}$. 

The theoretical value of $\theta_{13}$ is obtained as follows. Since $V_{ud} \sim 1$, $V_{us} \sim \lambda$ and $V_{ub} \sim \lambda^3$, where $\lambda = \sin \theta_{\rm C} \approx 0.2253$~\cite{Agashe:2014kda}, eq.~\eqref{eq:CKM-PMNS} can be written as 
\begin{align}
	U_{e3} = - \lambda U_{\mu 3} + {\cal O}(\lambda^2). \label{eq:CKM-PMNS_lowest}
\end{align}
This is the lowest order approximation of eq.~\eqref{eq:CKM-PMNS}. From eq.~\eqref{eq:CKM-PMNS_lowest}, the following is obtained,
		\begin{align}
			\tan^2 \theta_{13} = \sin^2 \theta_{\rm C} \sin^2 \theta_{23} + {\cal O}(\lambda^3).
			\label{eq:tan2theta_13}
		\end{align}
		This is the relation among three mixing angles---two lepton mixing angles, $\theta_{13}$, $\theta_{23}$, and the Cabibbo angle, $\theta_{\rm C}$. Eq.~\eqref{eq:tan2theta_13} shows that larger (smaller) $\theta_{23}$ gives larger (smaller) $\theta_{13}$. In the limit of the maximal 2-3 mixing,  eq.~\eqref{eq:tan2theta_13} gives $\theta_{13} \simeq \theta_{\rm C}/\sqrt{2}$. A factor $1/\sqrt{2}$ is originated from $\sin \theta_{23}$. 
		
		Using the global data of $\theta_{23}$~\cite{Gonzalez-Garcia:2014bfa}, $\theta_{23}=42.3{_{-1.6}^{+3.0}}^\circ$ for normal mass ordering, $\theta_{23}=49.5{_{-2.2}^{+1.5}}^\circ$ for inverted mass ordering, the theoretical value of $\theta_{13}$ is obtained as
		\begin{align}
			\theta_{13} = 
				\begin{cases}
					8.6{_{-0.3}^{+0.5}}^\circ & (\mbox{Normal mass ordering}), \\
					9.7{_{-0.3}^{+0.2}}^\circ & (\mbox{Inverted mass ordering}),
				\end{cases}
				\label{eq:theta_13}
		\end{align}
		where the best fit value $\pm 1\sigma$ is shown. As shown in eq.~\eqref{eq:theta_13}, in the case of normal mass ordering, the theoretical value of $\theta_{13}$ is in good agreement with the current global data $\theta_{13}=8.50{_{-0.21}^{+0.20}}^\circ$~\cite{Gonzalez-Garcia:2014bfa}. In the case of inverted mass ordering, the theoretical value is far from the current global data $\theta_{13}=8.51{_{-0.21}^{+0.20}}^\circ$~\cite{Gonzalez-Garcia:2014bfa}.

Next we study the leptonic CP violation. The leptonic Dirac CP phase $\delta_{\rm CP}$ is shown in fig.~\ref{fig:triangle}. It shows that $\delta_{\rm CP} \sim \pi$. Thus, $|\sin\delta_{\rm CP}|$ is typically very small. Since $|{\rm Im}(V_{ub}U_{\tau3})| \sim {\cal O}(\lambda^3)$ and $|V_{us}U_{\mu 3}| \sim {\cal O}(\lambda)$, $|\sin\delta_{\rm CP}|$ should be of ${\cal O}(\lambda^2)$. 
	
The theoretical value of $\delta_{\rm CP}$ is obtained as follows. We write eq.~\eqref{eq:CKM-PMNS} as
\begin{align}
	\left( 1 - \frac{\lambda^2}{2}\right) U_{e3} + \lambda U_{\mu 3} 
	+ A\lambda^3 (\overline{\rho} - i \overline{\eta}) U_{\tau 3} = 0 + {\cal O}(\lambda^4),
	\label{eq:CKM-PMNS_2nd}
\end{align}
where $A$, $\lambda$, $\overline{\rho}$ and $\overline{\eta}$ are the usual Wolfenstein parameters~\cite{Wolfenstein:1983yz, Agashe:2014kda}. From eq.~\eqref{eq:CKM-PMNS_2nd}, we obtain
\begin{align}
	\sin \delta_{\rm CP} = - A \lambda^2 \overline{\eta} \cot\theta_{23},
	\label{eq:sinCP}
\end{align}
and $\cos\delta_{\rm CP} \simeq -1$. Thus, in the case that eq.~\eqref{eq:nu_e_nu3} is exact, the magnitude of $|\sin\delta_{\rm CP}|$ is very small, which is consistent with the results in refs.~\cite{Dasgupta:2014ula, Girardi:2015vha}. Using the global data of the Wolfenstein parameters~\cite{Agashe:2014kda}, the leptonic Dirac CP phase is obtained as $\delta_{\rm CP} = (180.6\mbox{ - }181.1)^\circ$. It is too small to be observed by neutrino oscillation experiments. However, an interesting possibility has been reported in ref.~\cite{Razzaque:2014vba}---even if $\delta_{\rm CP}$ is close to $\pi$, the leptonic CP violation can be observed in atmospheric neutrinos. 

Here we consider the deviation from eq.~\eqref{eq:nu_e_nu3}. The deviation appears in the r.h.s in eq.~\eqref{eq:CKM-PMNS}. Since it represents the mixing between $\nu_e^\prime$ and $\nu_3$, we write it as $U_{e3}^\prime$. In general, $U_{e3}^\prime$ is complex. In this case, eq.~\eqref{eq:CKM-PMNS} can be replaced by 
\begin{align}
	V_{ud} U_{e3} + V_{us} U_{\mu 3} + V_{ub} U_{\tau 3} = U_{e3}^\prime.  \label{ep:CKM-PMNS2}
\end{align}
		\begin{figure}[h]
			\begin{center}
				\includegraphics[width=140mm]{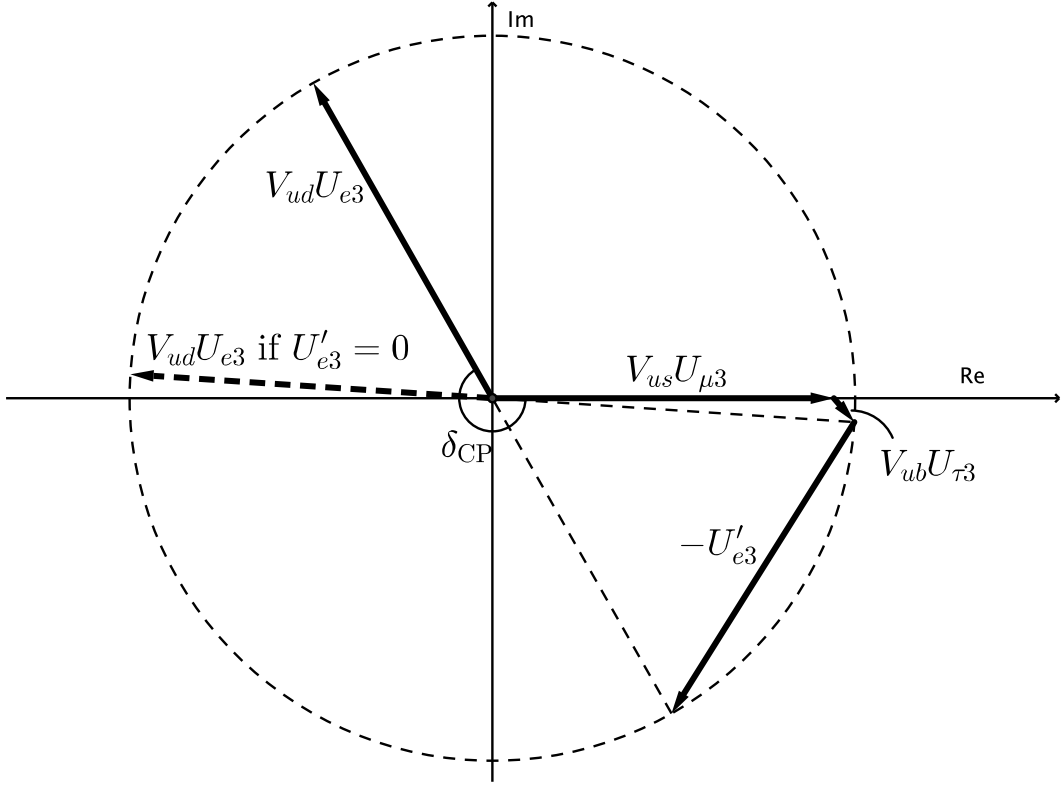}
				\caption{\label{fig:triangle2}Schematic figure of eq.~\eqref{ep:CKM-PMNS2}. The figure shows the special case that $|U_{e3}^\prime|$ is large. }
			\end{center}
		\end{figure}
Fig.~\ref{fig:triangle2} is a schematic figure of eq.~\eqref{ep:CKM-PMNS2}. In fig.~\ref{fig:triangle2}, the length of radius of the circle equals to $|V_{ud}U_{e3}| \simeq \sin \theta_{13}$. Fig.~\ref{fig:triangle2} shows the special case that $|U_{e3}^\prime|$ is large. In the case that $|U_{e3}^\prime|$ is small, fig.~\ref{fig:triangle2} is essentially equivalent to fig.~\ref{fig:triangle}. 

As shown in fig.~\ref{fig:triangle2}, by tuning both the magnitude and the argument of $U_{e3}^\prime$ appropriately, any values of the leptonic CP phase $\delta_{\rm CP}$ can be obtained, including $\delta_{\rm CP}\sim 3\pi/2$, with keeping the successful prediction $\theta_{13} \simeq 9^\circ$. However, theoretically, this tuning is unnatural in the following meaning. In the case that $|\sin\delta_{\rm CP}|$ is large, the value of $|V_{ud}U_{e3}| \simeq \sin \theta_{13}$ depends on $|U_{e3}^\prime|$. Therefore, in this case, $\theta_{13} \simeq 9^\circ$ is originated from tuning of $|U_{e3}^\prime|$. It is artificial. In spite of the best fit value $\theta_{13}=0^\circ$ at that time~\cite{Maltoni:2004ei}, the reason why some theorists have believed $\theta_{13} \simeq 9^\circ$ before the measurements is that $\sin\theta_{13}$ is essentially determined only by $|V_{us}U_{\mu3}|$. In that prediction there is no tuning. From these investigations, we conclude as follows---any values of $\delta_{\rm CP}$ can be obtained by fine tuning. However, without fine tuning, $|\sin\delta_{\rm CP}|$ should be very small, typically $|\sin\delta_{\rm CP}| \sim {\cal O}(\lambda^2)$. 

Finally, we propose a measure of CP violation. We consider,
\begin{align} 
	\frac{U_{e1} U_{e2} U_{e3}^* U_{\mu 3} U_{\tau 3}}{1-|U_{e3}|^2}. \label{eq:leptonicCP} 
\end{align}
It should be noted that we do not consider Majorana CP phases in this paper, and therefore the matrix elements in eq.~\eqref{eq:leptonicCP} should be understood as those of $V_{\rm PMNS}$, where $U_{\rm PMNS}=V_{\rm PMNS} P$ with $P$ the diagonal Majorana phase matrix. Eq.~\eqref{eq:leptonicCP} is {\it not} invariant under the transformation $U_{\alpha i} \rightarrow e^{i\varphi_\alpha} U_{\alpha i} e^{-i\varphi_i}$, where $\varphi_\alpha (\alpha=e, \mu, \tau)$ and $\varphi_i (i=1, 2, 3)$ are arbitrary. However, eq.~\eqref{eq:leptonicCP} {\it is} invariant under this transformation if $\varphi_e + \varphi_\mu + \varphi_\tau= 0$ and $\varphi_1 + \varphi_2 + \varphi_3 = 0$. Since the overall phase does not change the physics, without loss of generality we can impose these conditions. Therefore, it is concluded that eq.~\eqref{eq:leptonicCP} is phase-convention independent, similarly to the usual Jarlskog parameter $J_{\rm CP}$~\cite{Jarlskog:1985ht}. 

In terms of the standard parametrization~\cite{Agashe:2014kda}, we find that 
\begin{align}
	J_{\rm CP}
	=
	{\rm Im}\frac{ U_{e1} U_{e2} U_{e3}^* U_{\mu 3} U_{\tau 3} }{1-|U_{e3}|^2}
	&= 
	\frac{1}{8}\sin2\theta_{12} \sin2\theta_{23} \sin2\theta_{13} \cos\theta_{13}
	\sin \delta_{\rm CP}, \label{eq:Im_leptonicCP} \\
	\overline{J}_{\rm CP}
	\equiv 
	{\rm Re }\frac{ U_{e1} U_{e2} U_{e3}^* U_{\mu 3} U_{\tau 3} }{1-|U_{e3}|^2}
	&= 
	\frac{1}{8}\sin2\theta_{12} \sin2\theta_{23} \sin2\theta_{13} \cos\theta_{13}
	\cos \delta_{\rm CP}. \label{eq:Re_leptonicCP}
\end{align}
Thus, the imaginary part of eq.~\eqref{eq:leptonicCP} equals to the usual Jarlskog parameter $J_{\rm CP}$, and the real part gives $\overline{J}_{\rm CP}$, which is proportional to $\cos\delta_{\rm CP}$. The coefficients of $J_{\rm CP}$ and $\overline{J}_{\rm CP}$ are identical. Since eq.~\eqref{eq:leptonicCP} is phase-convention independent, the scheme of using eq.~\eqref{eq:leptonicCP} is applicable to any flavor models. 

To show an example, we apply this scheme to the quark sector. We consider 
\begin{align}
	\frac{V_{ud} V_{us} V_{ub}^* V_{cb} V_{tb}}{1 - |V_{ub}|^2}, \label{eq:quarkCP}
\end{align}
which is phase-convention independent. Using the Wolfenstein parametrization~\cite{Wolfenstein:1983yz, Agashe:2014kda}, the imaginary part and the real part of eq.~\eqref{eq:quarkCP} are given by 
\begin{align}
	J_{\rm CP}^{\rm (quark)} &= {\rm Im} \frac{V_{ud} V_{us} V_{ub}^* V_{cb} V_{tb}}{1 - |V_{ub}|^2} = A^2 \lambda^6 \overline{\eta}, \\
	\overline{J}_{\rm CP}^{\rm (quark)} &=  {\rm Re} \frac{V_{ud} V_{us} V_{ub}^* V_{cb} V_{tb}}{1 - |V_{ub}|^2} = A^2 \lambda^6 \overline{\rho}. 
\end{align}	
Their ratio equals to $\tan \delta_{\rm CP}^{\rm (quark)}$, 
\begin{align}
	\frac{J_{\rm CP}^{\rm (quark)}}{\overline{J}_{\rm CP}^{\rm (quark)}} = \tan \delta_{\rm CP}^{\rm (quark)} 	= {\overline{\eta}}/{\overline{\rho}}= 2.64. 
\end{align}
	Therefore, the Kobayashi-Maskawa CP phase in the quark sector is given by $\delta_{\rm CP}^{\rm (quark)} \simeq 69^\circ$.

In summary, we have considered a minimal condition that predicts $\theta_{13} \simeq \theta_{\rm C}/\sqrt{2}$, and have presented the improved prediction of $\theta_{13}$. The theoretical value of $\theta_{13}$ has been given in eq.~\eqref{eq:theta_13}. In the case of normal mass ordering, the predicted value is in good agreement with the current global best fit. In the case of inverted mass ordering, the predicted value is far from the current global best fit. We have also shown that any values of the leptonic Dirac CP phase $\delta_{\rm CP}$ can be obtained by fine tuning, including $\delta_{\rm CP} \simeq 3\pi/2$. Without fine tuning, $|\sin\delta_{\rm CP}|$ should be very small, typically of ${\cal O}(\lambda^2)$. Furthermore, we have proposed a measure of CP violation, eq.~\eqref{eq:leptonicCP}, which is applicable to any flavor models. If our predictions, eqs.~\eqref{eq:tan2theta_13} and~\eqref{eq:sinCP}, will be confirmed by future experiments, eq.~\eqref{eq:nu_e_nu3} should be taken seriously.

\bibliography{references}

\end{document}